# Context-free ordinals

Z. Ésik and S. Iván
Dept. of Computer Science
University of Szeged
Szeged, Hungary

November 21, 2018

## Abstract

We consider context-free languages equipped with the lexicographic ordering. We show that when the lexicographic ordering of a context-free language is scattered, then its Hausdorff rank is less than $\omega^\omega$. As a corollary of this result we obtain that an ordinal is the order type of a well-ordered context-free language iff it is less than $\omega^{\omega^\omega}$.

## 1 Introduction

When the alphabet $\Sigma$ of a language $L \subseteq \Sigma^*$ is linearly ordered, we may linearly order $L$ with the lexicographic order $<_{\mathrm{lex}}$. We call $L$ well-ordered, scattered, or dense when $(L, <_{\mathrm{lex}})$ has the appropriate property.

Efficient algorithms exist to decide whether or not a regular language (given by a deterministic or nondeterministic finite automaton) is scattered or a well-ordering, cf. [3, 11]. It is well-known that an ordinal is the order type of a well-ordered regular language iff it is less than $\omega^\omega$. Moreover, the Hausdorff rank of a scattered regular language is less than $\omega$, cf. [2, 14, 15].

The study of the lexicographic orderings of context-free languages was initiated in [4]. It is decidable for a context-free grammar whether it generates a well-ordered of scattered language [13]. In contrast, it is undecidable for a context-free grammar whether the language generated by it is dense, cf. [12]. Call an ordinal context-free if it is the order type of a well-ordered context-free language. In [4, 5], it was shown that every ordinal less than $\omega^{\omega^\omega}$ is a context-free ordinal and it was conjectured that no other ordinals are context-free. In this note we confirm this conjecture. Moreover, we show that the Hausdorff rank of a scattered context-free language is less than $\omega^\omega$. These facts were formerly known only for deterministic context-free languages and languages generated by prefix grammars [5, 6].

## 2 Linear orderings

A linear ordering is a pair $(P, <)$ where $P$ is some set and $<$ is a transitive binary relation on $P$ such that for each $x, y \in P$, exactly one of $x < y$, $y < x$ and $x = y$ holds. We will sometimes denote a linear ordering $(P, <)$ by just $P$. When $P_1 = (P_1, <_1)$ and $P_2 = (P_2, <_2)$ are linear orderings, a function $h : P_1 \to P_2$ is an embedding of $P_1$ into $P_2$ if $h(x) <_2 h(y)$ for each $x, y \in P_1$ with $x <_1 y$. If $h$ is also surjective, $h$ is an isomorphism. We call an isomorphism class an *order type*.

Examples of linear orderings include the finite linear orderings and the ordering $\mathbb{Z}$ of the integers, ordered as usual.

The *ordered sum* $P_1 + P_2$ of linear orderings $P_1, P_2$, or more generally, the ordered sum $\sum_{x \in Q} P_x$, where $Q$ is any linear ordering and for each $x \in Q$, $P_x$ is a linear ordering, are defined as usual, see e.g. [16]. The sum operation may be extended to order types. Suppose that $(P, <)$ is a linear ordering and that $P$ is the union of its subsets $Q_1$ and $Q_2$. Then $(Q_1, <)$ and $(Q_2, <)$ are linear orderings, and we call $(P, <)$ the *union* of $(Q_1, <)$ and $(Q_2, <)$. When in addition $Q_1$ and $Q_2$ are disjoint, then $(P, <)$, or any linear ordering isomorphic to $(P, <)$ is called a *shuffle* of $(Q_1, <)$ and $(Q_2, <)$.

A linear ordering $(P, <)$ is a *well-ordering* if there is no infinite descending chain $x_1 > x_2 > \ldots$ in $P$. An *ordinal* is the order type of a well-ordering. It is known that any set of ordinals is well-ordered by the relation $\alpha < \beta$ if and only if $\alpha \neq \beta$ and some well-ordering of order type $\alpha$ can be embedded into a well-ordering of order type $\beta$ iff there is some nonzero ordinal $\gamma$ with $\alpha + \gamma = \beta$.

A linear ordering $(P, <)$ is a *dense* ordering if $P$ has at least two elements and for each $x, y \in P$, if $x < y$ then there exists some $z \in P$ with $x < z < y$. A linear ordering $(P, <)$ is *scattered* if no dense ordering can be embedded into it. It is clear that every well-ordering is scattered. It is well-known that every scattered sum of scattered linear orderings is scattered, and any well-ordered sum of well-orderings is a well-ordering. Moreover, any finite union or shuffle of scattered linear orderings is scattered, and any union or shuffle of well-orderings is a well-ordering. Moreover, if $P$ can be embedded into $Q$ and $Q$ is scattered or a well-ordering, then so is $P$.

Hausdorff classified the countable scattered linear orderings with respect to their rank. Our definition from [15] is a slight modification of the original. For each countable ordinal $\alpha$ we define the class $H_\alpha$ of countable linear orderings as follows. $H_0$ consists of all finite linear orderings, and when $\alpha > 0$ is a countable ordinal, then $H_\alpha$ is the least class of linear orderings closed under finite ordered sum which contains all linear orderings isomorphic to an ordered sum $\sum_{i \in \mathbb{Z}} P_i$, where each $P_i$ is in $H_{\beta_i}$ for some $\beta_i < \alpha$. By Hausdorff's theorem, a countable linear order $P$ is scattered iff it belongs to $H_\alpha$ for some countable ordinal $\alpha$. The *rank* $\mathrm{r}(P)$ of a countable linear ordering is the least ordinal $\alpha$ with $P \in H_\alpha$.

*From now on, all linear orderings will be assumed to be countable.* In the sequel we will use the following facts without mention.

**Fact 1** *If $P_1$ is a scattered linear ordering and $P_2$ embeds into $P_1$, then $\mathrm{r}(P_2) \leq \mathrm{r}(P_1)$.*

**Fact 2** *If $P_1$ and $P_2$ are scattered of rank $\alpha_1$ and $\alpha_2$, respectively, then the rank of the scattered linear ordering $P_1 + P_2$ is $\max\{\alpha_1, \alpha_2\}$. If $Q$ is scattered with $\mathrm{r}(Q) \leq 1$ and for each $x \in Q$, $P_x$ is scattered with $\mathrm{r}(P_x) < \alpha$, then the rank of the scattered linear ordering $\sum_{x \in Q} P_x$ is at most $\alpha$.*

**Fact 3** *If $\sum_{i \in \mathbb{Z}} P_i$ embeds into a scattered ordering $P$ and $\alpha$ is an ordinal such that $\mathrm{r}(P_i) \geq \alpha$ for infinitely many $i \in \mathbb{Z}$, then $\mathrm{r}(P) \geq \alpha + 1$.*

The first two facts are well-known. We believe that Fact 3 is also well-known, but we could not locate it in the literature. For completeness, we have spelled out a proof in the Appendix.

## 2.1 Lexicographic orderings

Let $\Sigma$ be an alphabet and let $\Sigma^*$ stand for the set of all finite words over $\Sigma$, $\varepsilon$ for the empty word, $|u|$ for the length of the word $u$, $u \cdot v$ or simply $uv$ for the concatenation of $u$ and $v$. A language is an arbitrary subset $L$ of $\Sigma^*$, the concatenation of the languages $K$ and $L$ is the language $K \cdot L = KL = \{uv :$

$u \in K, v \in L\}$. When $K = \{u\}$, for some word $u$, we will sometimes write $u \cdot L$ or just $uL$ for $\{u\}L$.

Suppose that $\Sigma$ is equipped with a linear order $<$. We define two partial orderings on $\Sigma^*$, the *prefix order* $<_{\mathrm{pr}}$ and the *strict order* $<_{\mathrm{s}}$. For any words $u, v \in \Sigma^*$, $u <_{\mathrm{pr}} v$ if and only if $v = uw$ for some nonempty $w \in \Sigma^*$, and $u <_{\mathrm{s}} v$ if and only if there exist words $w, u', v' \in \Sigma^*$ and letters $a < b$ in $\Sigma$ with $u = wau'$ and $v = wbv'$. Then the set $\Sigma^*$ of all words is linearly ordered by the *lexicographic order* $<_{\mathrm{lex}} = <_{\mathrm{pr}} \cup <_{\mathrm{s}}$. Thus, for any language $L \subseteq \Sigma^*$, $(L, <_{\mathrm{lex}})$ is a linearly ordered set, called the *lexicographic ordering of $L$*. It is known that every (countable) linear ordering is isomorphic to the linear ordering of a language over the *binary alphabet* $\{0, 1\}$.

We call the language $L$ well-ordered, scattered etc. if its lexicographic ordering has the appropriate property. When $L$ is scattered, we define define $\mathrm{r}(L)$ as $\mathrm{r}(L, <_{\mathrm{lex}})$. The order-type of a language $L$ is the order type of $(L, <_{\mathrm{lex}})$.

As mentioned in the Introduction, a *context-free ordinal* is any ordinal that is the order type of a well-ordered context-free language. For example, consider the binary alphabet $\{0, 1\}$, ordered by $0 < 1$. Then $0^*$ and $1^*0$ are well-ordered of order type $\omega$, the least infinite ordinal. For another example, consider the context-free language $\bigcup_{n \geq 0} 1^n 0(1^*0)^n$. It is well-ordered of order type $1 + \omega + \omega^2 + \ldots = \omega^\omega$. Thus, $\omega$ and $\omega^\omega$ are context-free ordinals. In Corollary 13 we will show that an ordinal is context-free iff it is less than $\omega^{\omega^\omega}$.

## 3 Union and shuffle

In this section, we give an estimate on the rank of the union or a shuffle of linear orderings.

A *tree domain* is a prefix-closed language in $\{0, 1\}^*$, i.e. a set $T \subseteq \{0, 1\}^*$ with $u \in T$, $v \leq_{\mathrm{pr}} u$ implying $v \in T$ for each $u, v \in \{0, 1\}^*$. (Hence if $T \neq \emptyset$, then $\varepsilon \in T$.) Words of $T$ are also called *nodes*. A *path* in a tree domain is a (possibly infinite) sequence $u_0 = \varepsilon, u_1, \ldots$ of nodes such that for each integer $n \geq 0$, if $u_{n+1}$ is defined then $u_{n+1} \in \{u_n 0, u_n 1\}$.

When $L \subseteq \{0, 1\}^*$ is a language and $u$ is a word, let $u^{-1}L$ stand for $\{v \in \{0, 1\}^* : uv \in L\}$. Clearly $u^{-1}L$ embeds into $L$, thus if $L$ is scattered, then so is $u^{-1}L$ with $\mathrm{r}(u^{-1}L) \leq \mathrm{r}(L)$. Moreover, if $W$ is a set of words that are pairwise incomparable with respect to the prefix order, then $\sum_{w \in W} w^{-1}L$ is isomorphic to $\bigcup_{w \in W} w(w^{-1}L)$ and thus embeds into $L$.

For each language $L \subseteq \{0,1\}^*$, let $\mathrm{Pref}(L)$ stand for the tree domain $\{v \in \{0,1\}^* : \exists u \in L, v \leq_{\mathrm{pr}} u\}$. (Equivalently, $\{v \in \{0,1\}^* : v^{-1}L \neq \emptyset\}$.)

When $T$ is a tree domain and $u \in \{0,1\}^*$, we also use the notation $T|_u$ for $u^{-1}T$, and refer to $T|_u$ as the *sub-tree domain* of $T$ rooted at $u$.

For an ordinal $\alpha$, let us denote by $\overline{\alpha}$ the (linearly ordered) set $\{\beta : \beta \leq \alpha\}$.[1] When $T$ is a tree domain and $\alpha$ is an ordinal, a *marking of $T$ over $\overline{\alpha}$* is a mapping $\varphi : \{0,1\}^* \to \overline{\alpha}$ satisfying the following conditions:

i) For any $u \in \{0,1\}^*$, $\varphi(u) = 0$ if and only if $T|_u$ is finite.

ii) For any $u \in \{0,1\}^*$, $\varphi(u) = \max\{\varphi(u \cdot i) : i \in \{0,1\}\}$.

iii) For any $u \in \{0,1\}^*$ with $\varphi(u) > 0$, the set
$$D_\varphi(u) = \{v \in \{0,1\}^* : \varphi(uv) = \varphi(u)\}$$
is a union of finitely many paths in $T|_u$.

By condition ii), for each $u \in \{0,1\}^*$ either $\varphi(u \cdot 0) = \varphi(u)$ or $\varphi(u \cdot 1) = \varphi(u)$. Thus, if $\varphi(u) > 0$ then $D_\varphi(u)$ is a union of a finite nonzero number of *infinite* paths.

The introduction of markings is motivated by the following fact:

**Proposition 4** *The following are equivalent for a language $L \subseteq \{0,1\}^*$ and ordinal $\alpha$:*

  i) *$L$ is scattered with $\mathrm{r}(L) \leq \alpha$;*

  ii) *there exists a marking $\varphi$ of $\mathrm{Pref}(L)$ over $\overline{\alpha}$;*

  iii) *$\mathrm{Pref}(L)$ is scattered with $\mathrm{r}(\mathrm{Pref}(L)) \leq \alpha$.*

**Proof.** The third condition clearly implies the first, since $L$ embeds into $\mathrm{Pref}(L)$.

To show i)→ii) let $L \subseteq \{0,1\}^*$ be a scattered language with $\mathrm{r}(L) \leq \alpha$ and let $T$ stand for $\mathrm{Pref}(L)$. We define $\varphi : \{0,1\}^* \to \overline{\alpha}$ by $\varphi(u) = \mathrm{r}(u^{-1}L)$, for all $u \in \{0,1\}^*$. Since $u^{-1}L$ embeds into $L$, we have $\varphi(u) \leq \alpha$ for each $u$.

---

[1] Of course, $\overline{\alpha}$ may be identified with the ordinal $\alpha + 1$.

Note that if $\varphi(u) > 0$, then $u \in T$ and $T|_u$ is infinite. If $\varphi(u) = 0$, then by the definition of the rank we have that $u^{-1}L$ is finite, thus $T|_u$ is finite as well.

Since for all words $u \in \{0,1\}^*$ we have $u^{-1}L - \{\varepsilon\} = (0 \cdot (u0)^{-1}L) \cup (1 \cdot (u1)^{-1}L)$ which is isomorphic to $(u0)^{-1}L + (u1)^{-1}L$, we have that $\varphi(u) = \max\{\varphi(u \cdot b) : b \in \{0,1\}\}$. It follows now that if $\varphi(u) > 0$, then $D_\varphi(u)$ is a union of some infinite paths (and is thus a tree domain).

Now assume that there exists some $u \in T$ with $\varphi(u) = \beta > 0$ such that the set $D_\varphi(u)$ is not a union of finitely many paths. Then $D_\varphi(u)$ is the union of an infinite number of infinite paths, so that there exists an infinite set $W$ of nodes in $D_\varphi(u)$ which are pairwise incomparable with respect to prefix order and such that $\varphi(uw) = \beta$ for each $w \in W$. Hence the ordered sum $\sum_{w \in W}(uw)^{-1}L$ that is isomorphic to the lexicographic ordering of $\bigcup_{w \in W} w(w^{-1}u^{-1}L)$ can be embedded into $u^{-1}L$, which is a contradiction, since the rank of each language $(uw)^{-1}L$ with $w \in W$ is $\beta$, thus $\mathrm{r}(u^{-1}L) > \beta$, finishing the proof of i)→ii).

For ii)→iii) let us write again $T$ for $\mathrm{Pref}(L)$ and let $\varphi : \{0,1\}^* \to \overline{\alpha}$ be a marking of $T$. We show by induction on $\varphi(u)$ that $T|_u$ is scattered and $\mathrm{r}(T|_u) \leq \varphi(u)$ for each node $u$ of $T$. When $\varphi(u) = 0$, by the definition of the marking $T|_u$ is finite, hence $T|_u$ is scattered with $\mathrm{r}(T|_u) = 0$.

Now let $\beta > 0$ and assume the claim holds for all $\gamma$ with $\gamma < \beta$. Since $\varphi$ is a marking of $T$, for each $u$ with $\varphi(u) = \beta$ we have that $D = D_\varphi(u)$ is a union of finitely many paths of $T|_u$. Consider the set of nodes $\widehat{D} = D \cup D0 \cup D1$. Then $\widehat{D}$, equipped with the lexicographic order, is scattered of rank 1. To see this, we use induction on the number $k$ of (infinite) paths covering D. If $k = 1$, then $D$ is a single infinite path and $\widehat{D}$ can be embedded into a linear ordering of order type $\omega + \omega^*$, where $\omega^*$ is the order type of the negative integers, ordered as usual. Thus, $\mathrm{r}(\widehat{D}) = 1$. Now in the induction step, suppose that $k > 1$ and let $u \in \{0,1\}^*$ be the longest common prefix of the $k$ infinite paths covering $D$. Then let $D_i$ be the set of all nodes of $D$ of the form $uiv$, for $i = 0, 1$. Note that both $D_0$ and $D_1$ are finite unions of less than $k$ infinite paths. Also, $\widehat{D}$ is isomorphic to a sum $F_0 + \widehat{D}_0 + \widehat{D}_1 + F_1$, where $F_0$ and $F_1$ are finite. Since by the induction hypothesis $\widehat{D}_0$ and $\widehat{D}_1$ have rank 1, the same holds for $\widehat{D}$.

Then $\widehat{D}$ can be written as the union $\{v0 : v <_{\mathrm{pr}} u, v0 \not<_{\mathrm{pr}} u\} \cup \{u\} \cup u \cdot u^{-1}(\cup_{i \in [k]: u0 \in D_i} D_i) \cup u \cdot u^{-1}(\cup_{i \in [k]: u1 \in D_i} D_i) \cup \{v1 : v <_{\mathrm{pr}} u, v1 \not<_{\mathrm{pr}} u\}$, such that the lexicographic ordering of $\widehat{D}$ is isomorphic to the ordered sum of these five languages. Since the first two and the last of these languages are finite and the other two are covered by less than $k$ infinite paths, applying the induction hypothesis the claim is proved.

For each $v \in D$, let $L_v = \{\varepsilon\}$, and for each $v \in \widehat{D} - D$, let $L_v = T|_v$. Then $T|_u$ is $\bigcup_{v \in \widehat{D}} v \cdot L_v$ which is isomorphic to the ordered sum $\sum_{v \in \widehat{D}} L_v$. Since $\mathrm{r}(L_v) < \beta$ for each $v \in \widehat{D}$ and since $\widehat{D}$ is scattered of rank 1, we have that $T|_u$ is scattered and $\mathrm{r}(T|_u) = \mathrm{r}(\sum_{v \in \widehat{D}} L_v) \leq \beta$.  (END OF PROOF.)

Using the notion of marking, the following fact can be easily deduced:

**Proposition 5** *Suppose $\varphi_i$ is a marking of the tree domain $T_i$, $i = 0, 1$. Then $\varphi(u) = \max\{\varphi_i(u) : i \in \{0, 1\}\}$ is a marking of $T_0 \cup T_1$.*

**Proof.** First note that $\varphi(u) = 0$ for some $u \in \{0,1\}^*$ iff $\varphi_i(u) = 0$ for $i \in \{0,1\}$ iff $T_i|_u$ is finite for $i = 1, 2$ iff $T_u$ is finite. It is clear that for any $u \in \{0,1\}^*$,

$$\begin{aligned}
\max_{b \in \{0,1\}} \varphi(u \cdot b) &= \max_{b \in \{0,1\}} \max_{i \in \{0,1\}} \varphi_i(u \cdot b) \\
&= \max_{i \in \{0,1\}} \max_{b \in \{0,1\}} \varphi_i(u \cdot b) \\
&= \max_{i \in \{0,1\}} \varphi_i(u) \\
&= \varphi(u).
\end{aligned}$$

Finally, consider an arbitrary $u \in \{0,1\}^*$ with $\varphi(u) = \alpha > 0$. We show that $D_\varphi(u)$ is a finite union of paths. It is clear that for any $v \in \{0,1\}^*$ and $i \in \{0,1\}$, $\varphi_i(uv) \leq \alpha$. Hence,

$$\begin{aligned}
D_\varphi(u) &= \{v \in \{0,1\}^* : \varphi(uv) = \alpha\} \\
&= \{v \in \{0,1\}^* : \varphi_0(uv) = \alpha\} \cup \{v \in \{0,1\}^* : \varphi_1(uv) = \alpha\},
\end{aligned}$$

and since both of these sets are a union of finitely many paths (if $\varphi_i(u) = \alpha$, then this statement comes from the fact that $\varphi_i$ is a marking, if $\varphi_i(u) < \alpha$, then the corresponding set is empty, which is again a union of finitely many paths), so is their union.  (END OF PROOF.)

**Corollary 6** *For an arbitrary (countable) scattered linear ordering $P$ that is the union of the scattered linear orderings $Q_1$ and $Q_2$, $\mathrm{r}(P) = \max\{\mathrm{r}(Q_1), \mathrm{r}(Q_2)\}$.*

**Corollary 7** *If the scattered linear ordering $P$ is a shuffle of the scattered linear orderings $Q_1$ and $Q_2$, then $\mathrm{r}(P) = \max\{\mathrm{r}(Q_1), \mathrm{r}(Q_2)\}$.*

For well-orderings, Corollary 7 follows from Theorem 1.38 in [17].

## 4 Concatenation

In this section our aim is to prove that for scattered languages $K, L \subseteq \Sigma^*$ the concatenation $KL$ is scattered with $\mathrm{r}(KL) \leq \mathrm{r}(L) + \mathrm{r}(K)$. Actually we prove an extension of this result.

**Theorem 8** *Let $K \subseteq \Sigma^*$ be scattered of rank $\alpha$. Suppose that for each $w \in K$, $L_w \subseteq \Sigma^*$ is scattered with $\mathrm{r}(L_w) \leq \beta$. Then the language*

$$L' = \bigcup_{w \in K} wL_w$$

*is scattered with $\mathrm{r}(L') \leq \beta + \alpha$.*

**Proof.** First note that it suffices to prove the Theorem in the case when $\Sigma$ is the binary alphabet $\{0, 1\}$, since if $\Sigma$ has more than 2 elements then we can replace $K$ by $h(K)$ and each $L_w$ by $h(L_w)$, where $h : \Sigma^* \to \{0,1\}^*$ is an injective homomorphism preserving the lexicographic order such that the words $h(a)$, $a \in \Sigma$ are of equal length. So let us suppose from now on that $\Sigma = \{0, 1\}$.

If $\alpha = 0$ then $K$ is finite and $L'$ is a finite union of scattered languages of rank at most $\beta$. Thus, by Corollary 6, $L'$ is scattered with $\mathrm{r}(L') \leq \beta = \beta + \alpha$.

We proceed by induction on $\alpha$. Suppose that $\alpha > 0$ (so that $K$ is infinite) and consider a marking $\varphi : \{0,1\}^* \to \overline{\alpha}$ of $K$ with $\varphi(\varepsilon) = \alpha$. Then $D = D_\varphi(\varepsilon)$ is a finite nonempty union of infinite paths that we denote by $\widehat{D}$. We can partition $\widehat{D} = D \cup D0 \cup D1$ into 3 sets:

$$\begin{aligned} D_0 &= D \\ D_\ell &= \{w0 : w0 \in \widehat{D},\ w0 \notin D\} \\ D_r &= \{w1 : w1 \in \widehat{D},\ w1 \notin D\}. \end{aligned}$$

(Note that if $w0 \in D_\ell$ then $w1 \in D$, and similarly, if $w1 \in D_r$, then $w0 \in D$.)

For each $w \in D_0$ let $L'_w = \{\varepsilon\}$ if $w \in L'$ and let $L'_w = \emptyset$ if $w \notin L$. Suppose now that $wi \in D_\ell \cup D_r$, where $i = 0, 1$. Then let

$$\begin{aligned} L'_{wi} &= \bigcup_{wiv \in K} v \cdot L_{wiv} \cup \bigcup_{uv = wi,\ u \in K,\ v \neq \varepsilon} v^{-1} L_u \\ &= \bigcup_{v \in (wi)^{-1} K} v \cdot L_{wiv} \cup \bigcup_{uv = wi,\ u \in K,\ v \neq \varepsilon} v^{-1} L_u. \end{aligned}$$

Note that $L'$ is isomorphic to

$$\sum_{w \in \widehat{D}} w \cdot L'_w.$$

Now since $\varphi(wi) < \alpha$, $(wi)^{-1}K$ is scattered of rank strictly less than $\alpha$. Thus, since for any word $v$ with $wiv \in K$ we have that $L_{wiv}$ is scattered of rank at most $\beta$, by the induction hypothesis we have that $\bigcup_{v \in (wi)^{-1}K} v \cdot L_{wiv}$ is scattered of rank less than $\beta + \alpha$. Also, for each $u \in K$ and $v \neq \varepsilon$ with $uv = wi$ we have that $L_u$ is scattered of rank at most $\beta$, so by Corollary 6, the finite union $\bigcup_{uv=wi,\ u \in K,\ v \neq \varepsilon} v^{-1} L_u$ is scattered of rank at most $\beta < \beta + \alpha$. (Recall that $wi$ is fixed.) Thus, by applying Corollary 6 again, we have that $L'_{wi}$ is scattered of rank strictly less than $\beta + \alpha$. Since $\widehat{D}$ is scattered of rank 1 and since $L'$ is isomorphic to $\sum_{w \in \widehat{D}} w \cdot L'_w$, it follows now that $L'$ is scattered of rank at most $\beta + \alpha$. (END OF PROOF.)

**Corollary 9** *If $K, L \subseteq \Sigma^*$ are scattered languages then $KL$ is scattered and $\mathrm{r}(KL) \leq \mathrm{r}(L) + \mathrm{r}(K)$.*

**Example 10** *Suppose that $\alpha, \beta$ are countable ordinals and let $K, L \subseteq \{0,1\}^*$ be well-ordered prefix languages of order type $\omega^\alpha$ and $\omega^\beta$, respectively. (Such laguages exist since every countable ordinal is the order type of a prefix language over $\{0,1\}$.) Then $KL$ is well-ordered of order type $\omega^\beta \times \omega^\alpha = \omega^{\beta+\alpha}$. Also, the Hausdorff ranks of $K$ and $L$ are $\alpha$ and $\beta$, and the rank of $KL$ is $\beta + \alpha$.*

## 5 Scattered context-free languages

A *context-free grammar* over the alphabet $\Sigma$ is a system $G = (N, \Sigma, P, S)$ where $N$ is the alphabet of nonterminals, $P$ is the finite set of productions and $S \in N$ is the start symbol. We use basic notions as usual. The *language $L(p)$ generated from a word* $p \in (N \cup \Sigma)^*$ is the set of all words $w \in \Sigma^*$ with $p \Rightarrow^* w$. The *context-free language $L(G)$ generated by $G$* is $L(S)$.

For any $X, Y \in N$, let $X \preceq Y$ if there exist some $p, q \in (N \cup \Sigma)^*$ with $X \Rightarrow^* pYq$. The *strong component* of a nonterminal $X$ consists of all nonterminals $Y$ such that $X \preceq Y$ and $Y \preceq X$. For strong components $\mathcal{C}$ and $\mathcal{C}'$, let $\mathcal{C} \preceq \mathcal{C}'$ if there exists $X \in \mathcal{C}$ and $Y \in \mathcal{C}'$ with $X \preceq Y$. When $\mathcal{C} \preceq \mathcal{C}'$ but $\mathcal{C} \neq \mathcal{C}'$ we write $\mathcal{C} \prec \mathcal{C}'$. The *height* of a strong component $\mathcal{C}$ is the largest integer $n$ such that there is a sequence $\mathcal{C}_0, \ldots, \mathcal{C}_n$ of strong components with $\mathcal{C}_n = \mathcal{C}$ and $\mathcal{C}_i \prec \mathcal{C}_{i+1}$ for all $i < n$. The height of a nonterminal is the height of its strong component.

The following fact was proved in [13].

**Theorem 11** *Suppose that $G = (N, \{0, 1\}, P, S)$ is a reduced context-free grammar which is $\varepsilon$-free and has no left recursive nonterminal. Then $L(G)$ is scattered iff for each strong component $\mathcal{C}$ containing a recursive nonterminal*[2] *there is a primitive word $u_0 = u_0^{\mathcal{C}}$, unique up to conjugacy, such that for all $X, Y \in \mathcal{C}$ there is a (necessarily unique) conjugate $v_0$ of $u_0$ and a proper prefix $v_1$ of $v_0$ such that if $X \Rightarrow^+ wYp$ for some $w \in \{0, 1\}^*$ and $p \in (N \cup \{0, 1\})^*$ then $w \in v_0^* v_1$.*

The above theorem is applicable for example for reduced context-free grammars in Greibach normal form. We use it to prove:

**Theorem 12** *The rank of every scattered context-free language is strictly less than $\omega^\omega$.*

**Proof.** First we note that it suffices to prove the theorem for nonempty context-free languages over the binary alphabet $\{0, 1\}$, not containing the empty word. Any such context-free language can be generated by a reduced context-free grammar in Greibach normal form. So suppose that $G = (N, \{0, 1\}, P, S)$ is a reduced context-free grammar in Greibach normal form generating the nonempty scattered language $L \subseteq \{0, 1\}^*$. We show that $\mathrm{r}(L) < \omega^\omega$.

Let $X$ be a nonterminal of height $h$. We prove the following fact.

*Claim.* Suppose that for each nonterminal $X'$ of height $h' < h$, $L(X')$ is scattered of rank at most $\omega^{h'} + 1$. Then $L(X)$ is scattered of height at most $\omega^h + 1$.

Suppose first that $X$ is not recursive. If $h = 0$ then $L(X)$ is finite and we are done. Suppose that $h > 0$. Then $L(X) = \bigcup \{L(p) : X \to p \in P\}$ and the height of each nonterminal occurring on the right side of any production $X \to p$ is strictly less than $h$. Thus, by Corollary 9, for each production $X \to p$, $L(p)$ is scattered of rank at most $\omega^{h-1} \times k(p)$ for some integer $k(p)$.

---
[2] A nonterminal $X$ is recursive if there exist $p, q \in (N \cup \{0, 1\})^*$ with $X \Rightarrow^+ pXq$.

Let $k = \max\{k(p) : X \to p \in P\}$. Then, by Corollary 6, $L(X)$ is scattered of rank at most $\omega^{h-1} \times k < \omega^h + 1$.

Suppose now that $X$ is recursive. Then let $u_0 = u_0^X$ and $u_\infty = u_0^\omega = u_0 u_0 \ldots$. Consider a finite prefix $u$ of $u_\infty$. Then exactly one of $u0$ and $u1$ is a prefix of $u_\infty$. Suppose that $u0$ is a prefix. Then consider all *left* derivations of the sort

$$X \Rightarrow^* wYp \Rightarrow u1q \tag{1}$$

where $Y \in N$, $p, q \in (N \cup \{0,1\})^*$, $w \in \{0,1\}^*$ such that $u1$ is not a prefix of $w$. There are a finite number of such derivations and for each such derivation each nonterminal occurring in $q$ is of height less than $h$. (Indeed, if for some derivation (1), $q$ contains a nonterminal $Z$ of height $h$, then $Z$ belongs to the strong component of $X$ and there exist words $v \in \{0,1\}^*$ and $r \in (N \cup \{0,1\})^*$ with $X \Rightarrow^* u1vZr$ which is a contradiction to Theorem 11.) Thus, by Corollary 9, for each $q$ there is an integer $k$ such that $\mathrm{r}(L(q)) \leq \omega^{h-1} \times k < \omega^h$. Let

$$L_u = \begin{cases} \bigcup_q 1L(q) & \text{if } u \notin L(X) \\ \{\varepsilon\} \cup \bigcup_q 1L(q) & \text{if } u \in L(X) \end{cases}$$

where $q$ is any word in a derivation (1). By Corollary 6, $L_u$ is scattered of rank less than $\omega^h$. When $u1$ is a prefix of $u_\infty$, define $L_u$ symmetrically.

We have that

$$L(X) = \bigcup_u u \cdot L_u$$

where $u$ ranges over all finite prefixes of $u_\infty$. Since the prefixes of $u_\infty$ form a scattered language of rank 1 and since for each prefix $u$, $L_u$ is scattered of rank less than $\omega^h$, by Theorem 8, $L(X)$ is scattered of rank at most $\omega^h + 1$.

Now by the above claim, it follows immediately by induction that when the height of $X$ is $h$, then $L(X)$ is scattered with $\mathrm{r}(L(X)) \leq \omega^h + 1$. Thus, $L = L(S)$ is scattered and $\mathrm{r}(L) < \omega^\omega$. (END OF PROOF.)

We say that an ordinal $\alpha$ is a *context-free ordinal* if there is a well-ordered context-free language $L$ whose order type is $\alpha$.

**Corollary 13** *An ordinal is context-free iff it is less than $\omega^{\omega^\omega}$.*

**Proof.** It is well-known that the Hausdorff rank of a well-ordering is less than $\omega^\omega$ iff its order type is less than $\omega^{\omega^\omega}$. On the other hand, every ordinal less than $\omega^{\omega^\omega}$ is context-free as shown in [4, 5]. (END OF PROOF.)

# 6 Conclusion

It was shown in [4] that any ordinal less than $\omega^{\omega^\omega}$ is a context-free ordinal. Moreover, it was proved in [5] that if $L$ is a well-ordered deterministic context-free language (or equivalently, $L$ is definable by an algebraic recursion scheme), or a well-ordered context-free language generated by a prefix grammar, then the order type of $L$ is less than $\omega^{\omega^\omega}$. However, the conjecture formulated in [5] that every context-free ordinal is less than $\omega^{\omega^\omega}$ remained open. In this note, we confirmed this conjecture. Interestingly, the same ordinals are definable by tree automata, cf. [10]. We have also shown that the Hausdorff rank of a scattered context-free language is less than $\omega^\omega$.

A hierarchy of recursion schemes and a corresponding hierarchy of grammars and language classes inside the Chomsky hierarchy were introduced in [8, 9]. These hierarchies are closely ralated to the Caucal hierarchy [7]. By extending results in [5, 6] and confirming some conjectures in [6], it was shown in [1] that an ordinal is definable by a recursion scheme of order $n$ iff it is less than $\omega \Uparrow (n+1) = \omega^{\cdot^{\cdot^{\cdot^\omega}}}$, a stack of $n+1$ $\omega$'s, and moreover, the rank of any scattered linear ordering definable by a scheme of order $n$ is less than $\omega \Uparrow n$.

We conjecture that an ordinal is the order type of the lexicographic ordering of a well-ordered language generated by a grammar of order $n$ iff it is less than $\omega \Uparrow (n+1)$. Moreover, we conjecture that the rank of any scattered language generated by a grammar of order $n$ is less than $\omega \Uparrow n$.

By Corollary 13 and the corresponding results in [5, 6], the context-free ordinals are exactly those ordinals that arise as order types of well-ordered deterministic context-free languages. However, it is not known whether there is a (scattered) context-free linear ordering which is not isomorphic to the lexicographic ordering of any deterministic context-free language. Since the monadic theory of any graph in the Caucal hierarchy is decidable, it follows that the monadic theory of the lexicographic ordering of any deterministic context-free language is decidable. Thus, if there is a context-free linear ordering with an undecidable monadic theory, then it follows that there is a context-free linear ordering that is not isomorphic to the lexicographic ordering of any deterministic context-free language.

# Appendix

In this Appendix, we provide a proof of Fact 3 that we were not able to locate in the literature.

We define for each ordinal $\alpha$ the following classes $V_\alpha$ and $F_\alpha$ of (countable) linear orderings, where $\alpha$ is a countable ordinal:

1. $V_0$ contains all the linear orderings having at most one element.

2. When $\alpha > 0$, $V_\alpha$ is the class of all orderings isomorphic to an ordered sum $\sum_{i \in \mathbb{Z}} P_i$ where each $P_i$ is in $V_{\alpha_i}$ for some $\alpha_i < \alpha$.

3. For each $\alpha \geq 0$, $F_\alpha$ is the class of all orderings isomorphic to a finite sum $P_1 + \ldots + P_n$ where each $P_i$ is in $V_\alpha$.

Now it is clear that $F_\alpha \subseteq H_\alpha$ and $V_\alpha \subseteq H_\alpha$ for all $\alpha$. We prove that also $H_\alpha \subseteq F_\alpha$ for all $\alpha$. This is clear when $\alpha = 0$. So suppose that $\alpha > 0$ and that our claim holds for all $\beta < \alpha$. Since $F_\alpha$ is closed under finite sums, in order to prove that $H_\alpha \subseteq F_\alpha$ it suffices to show that whenever $P$ is of the form $\sum_{i \in \mathbb{Z}} P_i$ with $P_i \in \bigcup_{\beta < \alpha} H_\beta$ for all $i$, then $P$ belongs to $F_\alpha$. But if $P_i \in H_{\beta_i}$, where $\beta_i < \alpha$, then by the induction hypothesis, also $P_i \in F_{\beta_i}$, thus $P_i$ is a finite sum of linear orderings in $V_{\beta_i} \subseteq H_{\beta_i}$. Thus, if $P = \sum_{i \in \mathbb{Z}} P_i$ with $P_i \in \bigcup_{\beta < \alpha} H_\beta$ for all $i \in \mathbb{Z}$, then $P = \sum_{i \in \mathbb{Z}} Q_i$ where $Q_i \in \bigcup_{\beta < \alpha} H_\beta$ for all $i \in \mathbb{Z}$, so that $P \in H_\alpha$.

We have shown that $F_\alpha = H_\alpha$ for all $\alpha$, so that for any scattered linear ordering $P$ and countable ordinal $\alpha$, $\mathrm{r}(P) \leq \alpha$ iff $P \in F_\alpha$.

In the rest of our argument, we will make use of Claim 1 and Claim 2:

*Claim 1. Suppose that $\{a\} + P + \{b\}$ embeds into $Q \in V_\alpha$ for some $\alpha > 0$. Then $\mathrm{r}(P) < \alpha$.*

Let $h$ be an embedding of $\{a\} + P + \{b\}$ into $Q$ and let us write $Q$ as $Q = \sum_{i \in \mathbb{Z}} Q_i$, where for each $i$, $Q_i$ is in $V_{\alpha_i}$ for some $\alpha_i < \alpha$. Then let $a'$ denote the unique integer with $h(a) \in Q_{a'}$, and similarly, let $b'$ denote the unique integer with $h(b) \in Q_{b'}$. Since $h$ is an embedding, it follows that $P$ embeds into $\sum_{a' \leq i \leq b'} Q_i$, which belongs to $F_\beta$ with $\beta = \max\{\alpha_i : a' \leq i \leq b'\}$. Thus, since each $\alpha_i$ is less than $\alpha$ we get $\mathrm{r}(P) \leq \beta < \alpha$.

*Claim 2. Suppose that $R$ is an infinite scattered linear ordering and for each $x \in R$, $P_x$ is a scattered with $\mathrm{r}(P_x) = \alpha > 0$. Then $\mathrm{r}(\sum_{x \in R} P_x) > \alpha$.*

Indeed, let $Q = \sum_{x \in R} P_x$ and suppose that $\mathrm{r}(Q) \leq \alpha$. Then $\mathrm{r}(Q) = \alpha$, so that $Q = Q_1 + \ldots + Q_n$ for some integer $n > 0$ and orderings $Q_i$ with $Q_i \in V_\alpha$ for all $i$. Since $R$ is infinite, there exist some $j = 1, \ldots, n$ and $x_1 < x_2 < x_3$ in $R$ such that $P_{x_1} + P_{x_2} + P_{x_3}$ embeds in $Q_j$ by some function $h$. Let $p_i \in P_{x_i}$ for $i = 1, 3$. Then $\{p_1\} + P_{x_2} + \{p_3\}$ embeds in $Q_j \in V_\alpha$, so that $\mathrm{r}(P_{x_2}) < \alpha$ by Claim 1, contrary to our assumptions. Thus $\mathrm{r}(Q) > \alpha$.

Now we are ready to complete the proof of Fact 3. Suppose that $\sum_{i \in \mathbb{Z}} P_i$ embeds into a scattered linear ordering $P$ and $\alpha$ is an ordinal such that $\mathrm{r}(P_i) \geq \alpha$ for all $i \in R$, where $R$ is an infinite subset of $\mathbb{Z}$. Then by Claim 2, $\sum_{i \in R} P_i$ is of rank at least $\alpha + 1$. Since $\sum_{i \in R} P_i$ embeds in $\sum_{i \in \mathbb{Z}} P_i$, the rank of $\sum_{i \in \mathbb{Z}} P_i$ is also at least $\alpha + 1$.